\documentclass[usenatbib]{mn2e}
\usepackage{psfig}

\def\ave#1{\langle #1 \rangle}

\def\simgr{\,\hbox{\hbox{$ > $}\kern -0.8em \lower
     1.0ex\hbox{$\sim$}}\,}
\def\simle{\,\hbox{\hbox{$ < $}\kern -0.8em \lower
     1.0ex\hbox{$\sim$}}\,}

\title[Non-stationary hyperaccretion of stellar-mass black
holes in three dimensions]{Non-stationary hyperaccretion of 
stellar-mass black holes in three dimensions: Torus evolution and
neutrino emission}
\author[S.~Setiawan, M.~Ruffert, and H.-Th.~Janka]{S.~Setiawan$^{1}$\thanks{E-mail:S.Setiawan@ed.ac.uk}, M.~Ruffert$^{1}$\thanks{E-mail:M.Ruffert@ed.ac.uk}, and H.-Th.~Janka$^{2}$\thanks{E-mail:thj@mpa-garching.mpg.de}\\
$^{1}$School of Mathematics, University of Edinburgh, Edinburgh, EH9 3JZ, 
Scotland, U.K.\\
$^{2}$Max-Planck-Institut f\"ur Astrophysik, Postfach 1317, 85741
Garching, Germany}
\begin{document}

\date{Accepted 0000. Received 0000; in original form 0000}

\maketitle

\begin{abstract}
We present three-dimensional hydrodynamic simulations of the
evolution of selfgravitating, thick accretion discs around 
hyperaccreting stellar-mass black holes. The black hole-torus 
systems are considered to be remnants of compact object
mergers, in which case the disc is not fed by an external mass
reservoir and the accretion is non-stationary.
Our models take into account viscous dissipation, 
described by an $\alpha$-law, a detailed equation of 
state for the disc gas, and an approximate treatment of general
relativistic effects on the disc structure by using a
pseudo-Newtonian potential for the black hole including 
its possible rotation and spin-up during accretion. 
Magnetic fields are ignored. The neutrino emission of the
hot disc is treated by a neutrino-trapping scheme, and 
the $\nu\bar\nu$-annihilation near the disc is evaluated
in a post-processing step. Our simulations show that the 
neutrino emission and energy deposition by $\nu\bar\nu$-annihilation
increase sensitively with the disc mass, with the black hole
spin in case of a disc in corotation, and in particular with
the $\alpha$-viscosity. We find that for sufficiently
large $\alpha$-viscosity $\nu\bar\nu$-annihilation can be a 
viable energy source for gamma-ray bursts.
\end{abstract}


\section{Introduction}

In a series of papers (Ruffert, Janka \& Sch\"afer 1996; 
Ruffert et al. 1997; 
Ruffert \& Janka 1999, 2001) it was shown that the neutrino emission
associated with the dynamical phase of the merging or collision 
of two neutron stars (NS+NS) is powerful, but too short to provide
the energy for gamma-ray bursts (GRBs) by neutrino-antineutrino annihilation. 
Significant heating of the coalescing stars occurs only after they
have plunged into each other. The neutrino luminosities can then
rise to several $10^{53}\,{\rm erg\,s}^{-1}$ (see also Rosswog \&
Liebend\"orfer 2003) and even exceed 
$10^{54}\,{\rm erg\,s}^{-1}$ in case of the more violent 
(although probably very rare) collisions. 

After a few milliseconds, the compact massive remnant of
the merger will most likely collapse to a black hole (BH)
(see, e.g., Popham, Woosley \& Fryer 1999, 
Di Matteo, Perna \& Narayan 2002, 
Oechslin et al. 2004, Shibata \& Ury\={u} 2000).
When this happens some matter can remain in a toroidal accretion disc
around the BH and a
funnel with low baryon density develops along the system axis.
Hydrodynamic simulations indicate that between some
0.01$\,M_{\odot}$ and a few 0.1$\,M_{\odot}$ of matter
may have enough angular momentum to resist immediate absorption
into the BH (see, e.g., Ruffert \& Janka 1999, Shibata \& Ury\={u} 
2000). A similar result was obtained for BH+NS mergers
(Janka et al.~1999, see also Lee 2001 and references therein). 
This matter is swallowed by the BH on the time scale of viscous
transport of angular momentum, which is much longer than
the dynamical time scale. Different from the collapsar case there
is no external reservoir of stellar matter which feeds the accretion
torus. Therefore the relevant time scale is set by viscosity-driven
accretion rather than infall. The accretion is only approximately 
stationary, if global instabilities do not play a role
(Ruffert et al.\ 1997, Lee \& Ramirez-Ruiz 2002).
Post-merger BH accretion
might explain short GRBs, but is unlikely to account for
long GRBs (e.g., Narayan, Piran \& Kumar 2001).

Post-merger accretion tori have maximum densities 
between some $10^{10}\,$g$\,$cm$^{-3}$ and 
more than $10^{12}\,$g$\,$cm$^{-3}$ and extend to outer radii
of 10--20 times the Schwarzschild radius ($R_{\mathrm{s}}
= 2GM/c^2$). The accretion proceeds with rates of
fractions of a solar mass per second up to several solar masses
per second. In case of such ``hyperaccreting'' black hole systems
(Popham et al.\ 1999) photons are strongly coupled to the torus 
plasma and therefore are inefficient in transporting away energy.
Neutrinos, however, are abundantly created by weak interactions 
in the very dense and hot tori and a fair fraction of the
gravitational binding energy of the accreted matter can be
radiated away by them (``neutrino-dominated accretion flow'', NDAF,
Popham et al.\ 1999, Ruffert et al.~1997, Ruffert \& Janka 
1999, Narayan et al.\ 2001, Kohri \& Mineshige 2002, 
Di Matteo et al.\ 2002).

In this Letter we present, for the first time, three-dimensional
(3D) simulations of time-dependent hyperaccretion from thick
tori on a stellar-mass BH including the physical effects of
on a stellar-mass BH including the physical effects of
BH rotation (Artemova, Bj\"ornsson \& Novikov 1996), 
viscosity ($\alpha$-prescription following Shakura \& Sunyaev 1973), 
a realistic finite-temperature equation of state 
according to Lattimer \& Swesty~\cite{lat91}, and energy loss
and change of lepton number by neutrino emission. The annihilation
of emitted neutrinos and antineutrinos is investigated for its
potential to provide the energy of
relativistic gamma-ray burst fireballs
(see, e.g., Eichler et al.\ 1989; Narayan, Paczy\'nski \& Piran 1992; 
Woosley 1993). Our 3D modeling of the time-dependent (non-stationary)
torus evolution abolishes approximations of previous (semi-)analytic
work (e.g., Popham et al.\ 1999, Di Matteo et al.\ 2002)
or more radically simplified numerical modeling in 2D
(Lee \& Ramirez-Ruiz 2002).

\section{Numerics and Initial Conditions}

We use an Eulerian, 3D hydrodynamics code 
based on the Piecewise Parabolic Method of Colella \&
Woodward (1984). The code employs three nested Cartesian grids 
in a computational volume of 500$\,$km side length.
Each coarser grid level has the same number of zones but twice
the zone size of the level below. Mirror symmetry relative to the
equatorial plane is assumed.
For an equatorial length and width of the computational volume of 
500$\,$km and a vertical extension of 125$\,$km,
the smallest zones have a side length of 1.95$\,$km. 
\begin{table*}
\caption[]{
Parameters and some results of the torus evolution models. In all
cases the initial BH mass was $M_{\rm BH}^{\rm i} = 4.017\,M_{\odot}$,
the number of zones of each level of the three nested grids was 64, 
and the simulation was performed over a time interval of 
$\Delta t_{\rm cal} = 40\,$ms.
$M_{\rm d}^{\rm i}$ and $a_{\rm i}$ are the gas mass and BH spin
parameter at the beginning of the simulations.
The direction of the BH rotation is indicated by ``pro'' and
``ret'', for prograde and retrograde relative to the disc,
respectively. The physical viscosity parameter is $\alpha$.
All the following quantities are given at the end of the simulations:
BH angular momentum parameter $a_{\rm f}$, torus mass $M_{\rm d}$,
BH mass accretion rate $\dot M_{\rm d}$,
typical accretion time scale of the torus, $t_{\rm acc}\equiv M_{\rm
d}/\dot M_{\rm d}$, average torus density $\ave{\rho_{\rm d}}$ and
average torus temperature $\ave{T_{\rm d}}$ for the bulk of the torus
gas, total neutrino luminosity, $L_\nu$, and integral rate
of energy deposition by $\nu\bar{\nu}$-annihilation around the
accretion torus, $\dot{E}_{\nu\bar{\nu}}$. The last column gives the 
total energy deposition by $\nu\bar\nu$-annihilation, 
$E_{\nu\bar{\nu}}$, in the time interval $\Delta t_{\rm cal}$.
}
\begin{flushleft}
\tabcolsep=2.0mm
\begin{tabular}{lcclcccccccccc}
\hline\\[-3mm]
model & 
   $M_{\rm d}^{\rm i}$ & BH & $\!\!\!$spin 
      & dir. & visc &
   $M_{\rm d}$ & $\dot M_{\rm d}$ & $t_{\rm acc}$ &
   $\ave{\rho_{\rm d}}$ & $\ave{T_{\rm d}}$ &
   $L_\nu$ & $\dot{E}_{\nu\bar{\nu}}$ & $E_{\nu\bar{\nu}}$
\\
 & 
   {\scriptsize $M_\odot$} & $a_{\rm i}$ & $\ \,a_{\rm f}$ & & $\alpha$ &
   {\scriptsize$10^{-2}M_\odot$} & {\scriptsize$M_\odot\,{\rm s}^{-1}$} & ms &
   {\scriptsize$10^{10}\frac{\rm g}{{\rm cm}^3}$} & MeV &
   {\scriptsize$10^{50}\frac{\rm erg}{\rm s}$} &
   {\scriptsize$10^{50}\frac{\rm erg}{\rm s}$} & 
   {\scriptsize$10^{50}{\rm erg}$}
\\[0.2ex] \hline\\[-3mm] 
r00-64  & 0.0478 & 0.0 & 0.0083 & --- & 0.0 & 2.22 & 0.29 & 77. & 1.0--1.5 & 1.5--2.0 & 1.8 & 1.8$\cdot 10^{-4}$  &  4.4$\cdot 10^{-4}$ \\  % 
ro2-64  & 0.0478 & 0.6 & 0.6023 &pro & 0.0 & 3.10 & 0.16 &194. & 1.0--1.5 & 2.0--2.5 & 3.4 &   7.2$\cdot 10^{-4}$  & 3.3$\cdot 10^{-3}$ \\  % p0.6
ro5-64  & 0.0478 & 0.6 & 0.5710 &ret & 0.0 & 1.13 & 0.25 & 45. & 0.4--0.5 & 1.5--2.0 &  1.0 &   3$\cdot 10^{-5}$  &  1.8$\cdot 10^{-4}$ \\  % r0.6   1 
ir1-64  & 0.0120 & 0.0 & 0.0006 & --- & 0.0 & 0.71 & 0.11 & 65. & 0.2--0.3 & 1.0      & 0.1 &   5$\cdot 10^{-7}$&  8$\cdot 10^{-5}$ \\  %  1/4
ir4-64  & 0.1912 & 0.0 & 0.0510 & --- & 0.0 & 7.87 & 1.28 & 61. & 2.0--3.0 & 2.5--3.5 & 22.4 & 2.8$\cdot 10^{-2}$  & 1.8$\cdot 10^{-3}$  \\  %    4
al3-64  & 0.0478 & 0.0 & 0.0079 & --- & 0.01& 2.44 & 0.25 & 98. & 1.0--1.5 & 3.0--4.0 & 11.4 &  8$\cdot 10^{-3}$  & 3.8$\cdot 10^{-3}$  \\    % al0.01
al4-64  & 0.0478 & 0.0 & 0.0082 & --- & 0.1 & 1.76 & 0.38 & 46. & 0.7--0.8 & 4.0--5.0 & 80.0  & 0.4  & 0.26   \\    % al0.1
ar1-64  & 0.0478 & 0.6 & 0.6024 &pro & 0.1 & 2.34 & 0.35 & 67. & 0.7--0.8 & 5.0--6.0 &110.0  & 0.7  & 0.29   \\    % al0.1 p0.6 1
ar2-64  & 0.1912 & 0.6 & 0.6331 &pro & 0.1 &10.44 & 0.28 &373. & 1.0--1.5 & 7.0--8.0 &440.0  & 11.0  &1.90     
  % al0.1 p0.6 4
     \\[0.2ex]
\hline
\end{tabular}
\end{flushleft}
\label{tab1}
\end{table*}

The gravitational potential of the black hole is described
by the Artemova-Bj\"ornsson-Novikov potential (Artemova et
al. 1996), which reduces to the Paczy\'nski-Wiita potential
(Paczy\'nski \& Wiita 1980) when the BH spin parameter
$a= Jc/(GM_{\rm BH}^2)$ is zero. 
This potential mimics some general relativistic effects
like the existence of an innermost stable circular orbit (ISCO) and
the variation of the ISCO and event horizon with the value of $a$.
The BH is represented as a gravitating ``vacuum sphere'' in 
our calculations. The corresponding inner boundary radius is defined by 
the arithmetic average of event horizon and ISCO. 
Selfgravity of the torus is taken into account by a Newtonian description. 
We use the $\alpha$-prescription proposed by 
Shakura \& Sunyaev~\cite{sha73} to model the physical effects of 
disc viscosity (in addition to the ever present numerical viscosity),
including the terms for viscous transport of angular momentum and
dissipation of energy 
(a detailed description of the technical implementation 
will be given elsewhere; Setiawan et al., in preparation). 
With the numerical resolution which we could
afford in the present set of simulations, the dissipative effects
of numerical viscosity should be somewhat larger than estimated
for our previous work (see Ruffert \& Janka 1999 and Janka et al. 1999)
and might correspond to an $\alpha$-viscosity around 0.01. The latter
value therefore sets a lower bound to the range of $\alpha$'s 
where we expect to observe influence by the inclusion of viscosity
terms in the simulations discussed here.
Gravitational-wave emission and its
corresponding back-reaction (Blanchet, Damour \& Sch\"afer 1990) are 
taken into account. A neutrino leakage scheme (Ruffert et
al.\ 1996) is used to calculate the energy loss and lepton number
change by neutrino emission from the disc gas.

The initial configurations in our simulations consist of a BH
surrounded by a toroidal accretion disc. These systems are 
considered as remnants of BH+NS mergers
and the system parameters were chosen appropriately. For this
purpose we 
were guided by data obtained in some of the model runs described 
in Janka et al. (1999). The simulations start
with a BH that has a mass of 4.017$\,M_\odot$ and an initial
rotational parameter of 0 or 0.6, respectively. Our 
``reference accretion torus'' has a mass 
$M_{\rm d}^{\rm i}$ of 0.0478$\,M_\odot$.
This mass was changed to 0.0120$\,M_\odot$ or 0.1912$\,M_\odot$
by simply scaling the density distribution with appropriate factors.
The initial temperature is around 2$\,$MeV in the inner
part of the torus (out to an equatorial distance from the system 
axis of $\sim\,$70$\,$km)
and still above 1$\,$MeV in the more dilute outer parts up to 
$\sim\,$130$\,$km. The different combinations of model parameters
of our simulations can be found in Table~\ref{tab1}.
Varing the system parameters by scaling the initial
torus density and setting the BH spin parameter might appear
ad hoc. One should notice, however, that a BH+NS merger is an 
extremely violent event which should not be expected to produce a 
BH-torus configuration in perfect rotational equilibrium. Moreover,
the torus properties will develop and adjust to the effects of
viscous shear. This initial phase of relaxation after the start
of our simulations does not last
very long and most transients have died out already after
$\sim 20\,$ms.

%*********************** fig01 : BH-acc, nulum, nubarnu *********************
\begin{figure*}
\tabcolsep=1.0mm
\begin{tabular}{lcr}
 \psfig{file=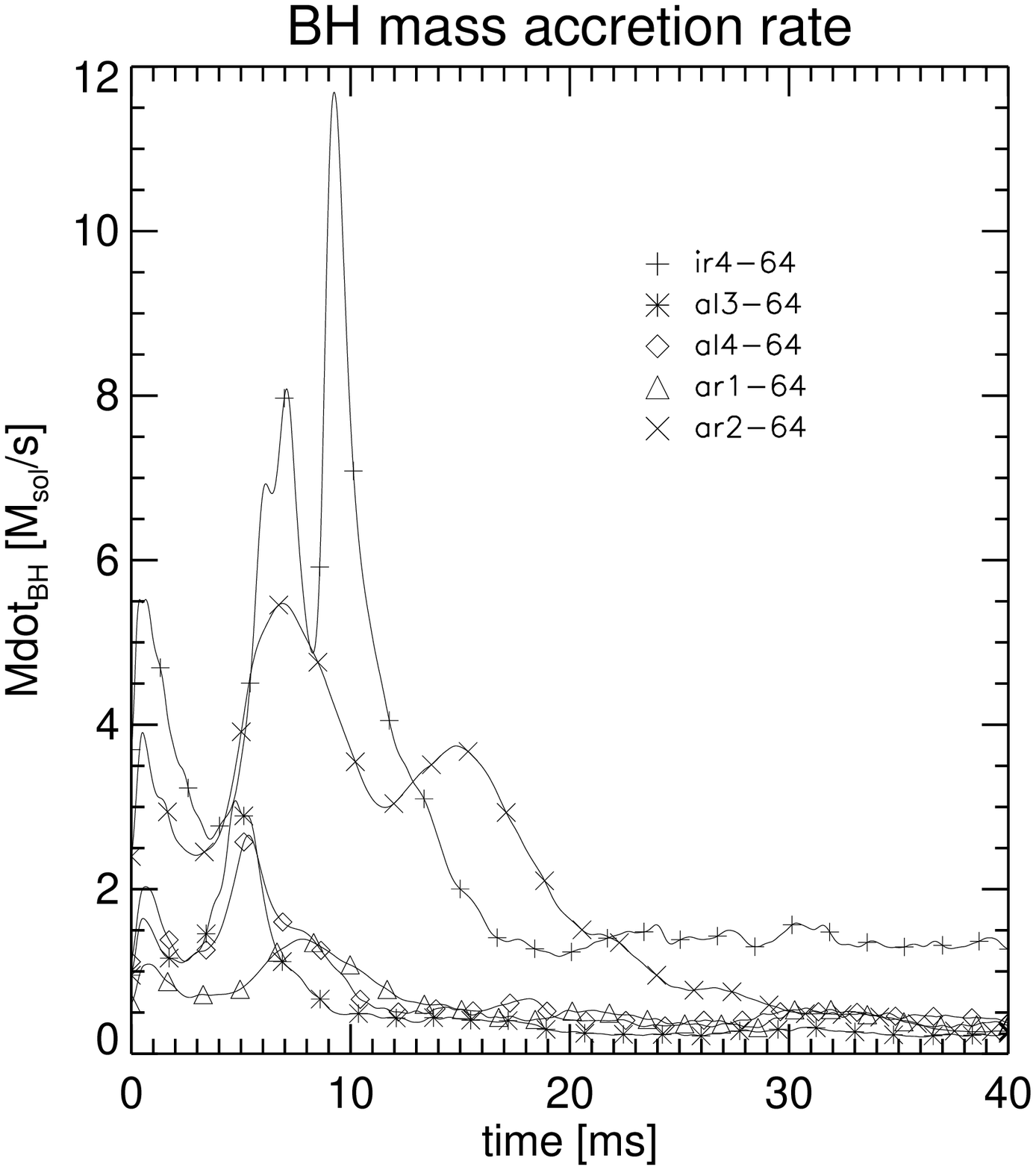,width=5.75cm,clip=} &
 \psfig{file=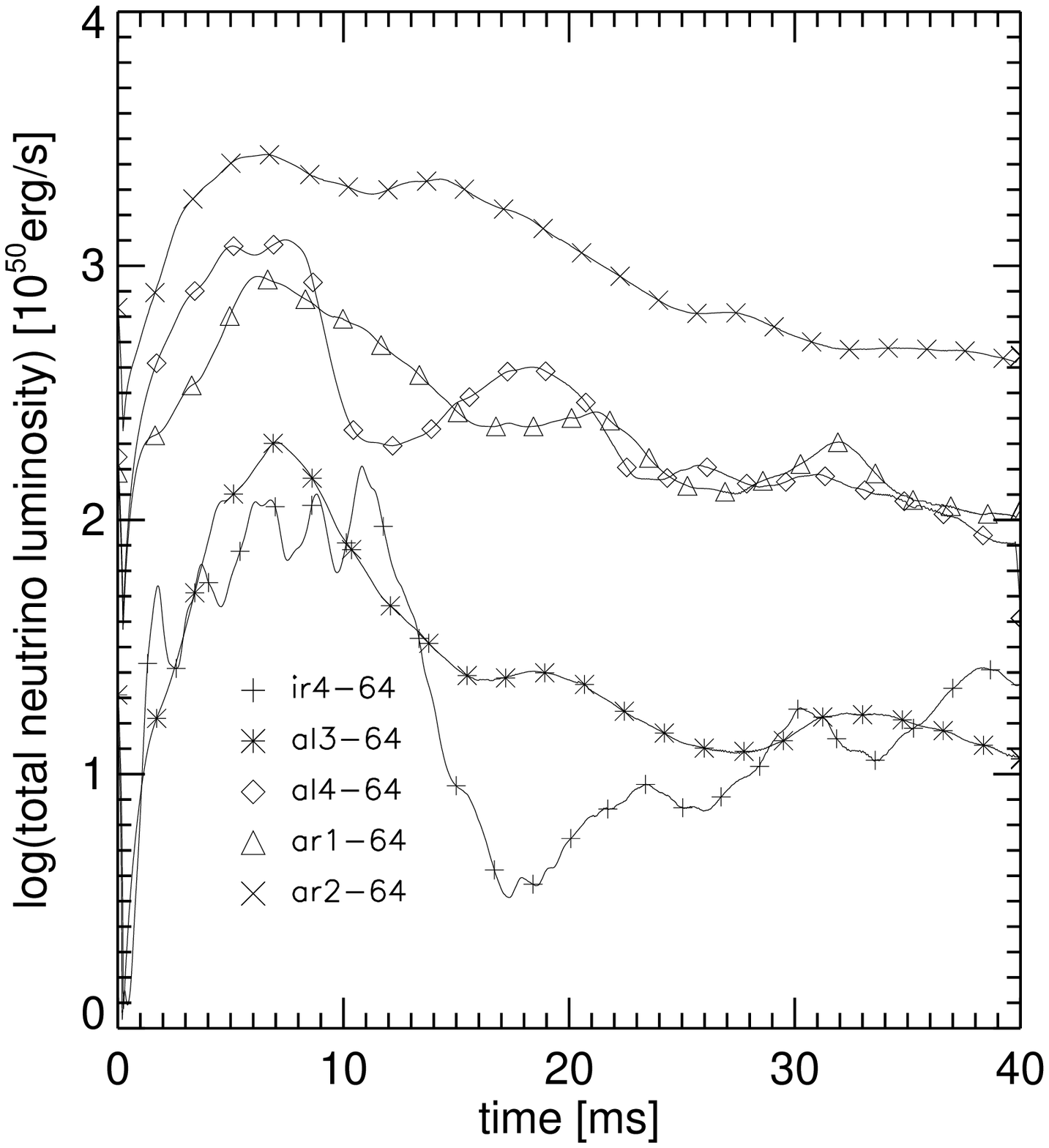,width=5.75cm,clip=} &
 \psfig{file=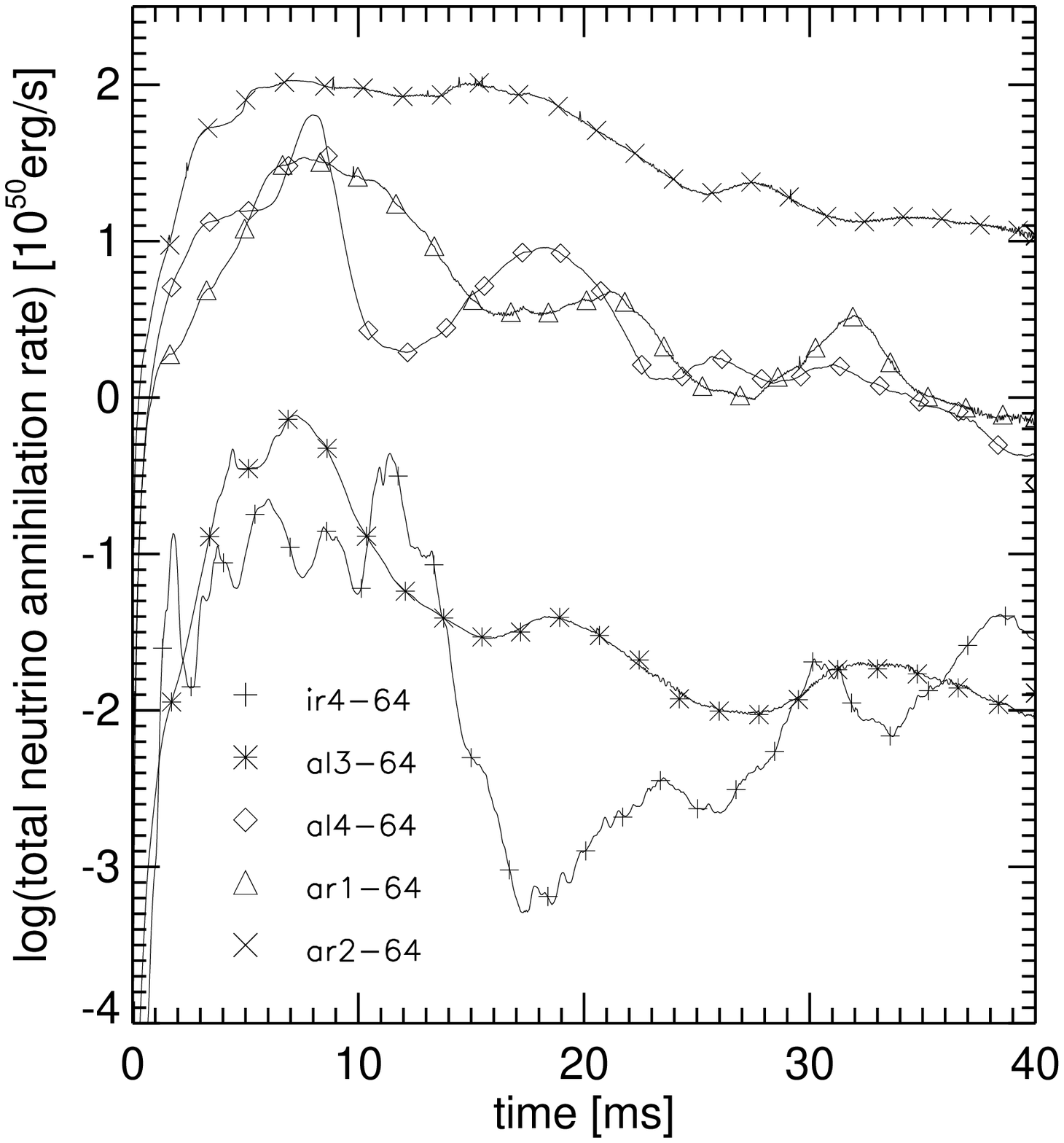,width=5.75cm,clip=}
\end{tabular}
\caption[]{Black hole mass accretion rate, $\dot{M}_{\rm d}$ (left), total 
neutrino luminosity, $L_{\nu}$ (middle), and integral rate of 
energy deposition by
$\nu\bar{\nu}$-annihilation, $\dot{E}_{\nu\bar{\nu}}$ (right),
as functions of time for different models. Note that only a small
fraction ($\sim\,1$\%) of this energy is released in the low-density
funnel above the poles of the BH. 
}
\label{fig01}
\end{figure*}
%*********************** fig01 : BH-acc, nulum, nubarnu *********************

\section{Results}

The models listed in Table~\ref{tab1} provide information about
the influence of the torus mass and viscosity, and of the BH rotation
on the system evolution.

The compact NDAFs considered here advect most of the disk mass into 
the BH, and convective currents do not play an important role 
for mass and angular momentum transport (cf.\ Narayan et al.\ 2001).
Angular momentum is carried outward mainly by shear effects. 
Because of the large ratio of BH to disk mass, the spin parameter
$a$ of the BH shows only minor variation in response to the angular 
momentum that is associated with the mass swallowed by the BH.
The value at the end of our computed evolution, $a_{\rm f}$, 
has increased for a corotating disc, otherwise has decreased
relative to $a_{\rm i}$ (see Table~\ref{tab1}). 

A comparison of Models r00-64 and al3-64, which differ only by
the presence of physical viscosity in the latter model,
confirms that $\alpha = 0.01$ is close to the threshold where
the effects of viscous terms become noticable. Except for the
temperature, which is increased by shear heating, both models
have very similar characteristic properties at the end of the
simulations. Model~al3-64 with the higher temperature has
a correspondingly larger neutrino emission and is somewhat more
inflated due to thermal pressure. The mass accretion rate is
therefore marginally lower and the remaining torus after
40$\,$ms is insignificantly more massive, in contrast to the 
consequences of a much higher viscosity (see below).

Concerning the simulated torus evolution general trends 
can be observed in dependence of the varied parameters (i.e.,
$M_{\rm d}^{\rm i}$, $a_{\rm i}$, and $\alpha$). The numbers
given in Table~\ref{tab1} reveal the following behaviour:
\begin{itemize}
\item  With higher disk mass $M_{\rm d}$
the mass accretion rate $\dot M_{\rm d}$ increases, 
but the accretion time scale, $t_{\rm acc}\equiv M_{\rm
d}/\dot M_{\rm d}$, shows little change. This is true, 
however, only for low disc viscosity and nonrotating BHs
(compare Models r00-64, ir1-64, ir4-64). In case of 
high torus viscosity and spinning BHs ($\alpha = 0.1$;
Models ar1-64 and ar2-64) a four times larger initial disk mass
correlates with a 5.6 times longer lifetime of the torus.
\item  A BH in corotation with the disc leads to a higher 
torus temperature, because the ISCO and event horizon shrink
and the torus moves closer to the BH. Since thermal pressure
puffs up the torus (as visible in Fig.~\ref{fig03}), its 
density does not rise in the same way.
The mass accretion rate decreases and the accretion time scale 
grows. Therefore the torus mass is higher at a given time.
Retrograde rotation of the BH has the opposite effect 
(compare Models r00-64, ro2-64, ro5-64 or al4-64 with ar1-64).
\item  Without physical viscosity, the tori remain relatively
cool (near their initial temperature), independent of the torus
mass. Larger viscosity (bigger $\alpha$) increases the mass
accretion rate and reduces the accretion time scale $t_{\rm acc}$.
This holds for nonrotating BHs (compare Models r00-64 and al4-64)
and rotating BHs (Models ro2-64 and ar1-64). With higher torus
viscosity the temperature becomes higher and the neutrino emission
correspondingly stronger. Torus mass and density are lower at 
the same time. The similarity of Models al4-64 and ar1-64, and
sizable differences relative to Model al3-64 suggest that 
increasing the viscosity from 0.01 to 0.1 is more significant
than BH rotation with $a = 0.6$ instead of no rotation.
\end{itemize}
The total neutrino luminosity of the torus, $L_{\nu}$, 
(i.e., the sum of the luminosities of neutrinos and antineutrinos 
of all flavors) increases with the torus mass,
with the BH spin in case of direct rotation (corotation with disc),
and with the viscosity. This is a consequence of a higher torus
temperature. Again the sensitivity is largest to a variation of
the viscosity within the considered bandwidth of values. 
For each model phases of high neutrino luminosity correlate with
large mass accretion rates (compare the left and middle panels of
Fig.~\ref{fig01}), indicating that the processes which lead
to a high value of $\dot{M}_{\rm d}$ (in particular during the
first 20$\,$ms of the computed evolution) also cause heating of
the torus plasma. But this correlation is not strict. This
can be seen in case of Model ir4-64 where after $\sim\,$15$\,$ms
the mass accretion rate becomes nearly constant but
$L_{\nu}$ shows a continuous trend of increase.

%******************************* fig02 *************************************
\begin{figure}
  \psfig{file=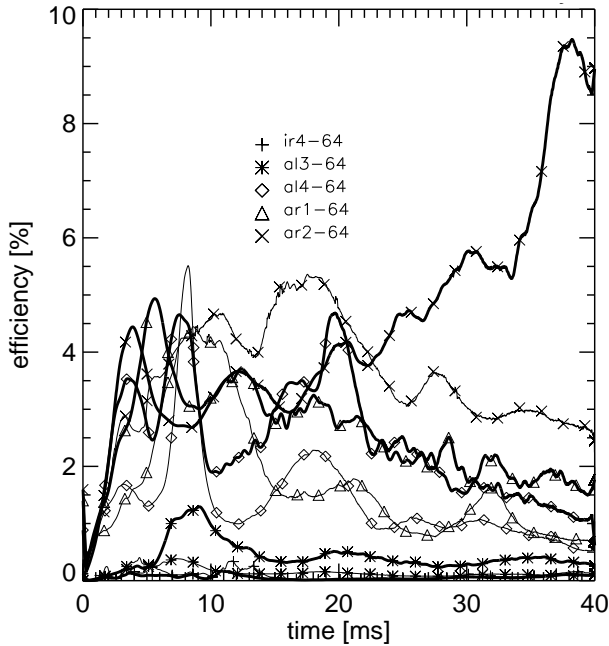,width=8.0cm,clip=}
\caption[]
{Efficiency of conversion of rest-mass energy of accreted matter
to neutrino emission, $q_{\nu}\equiv L_{\nu}/(\dot Mc^2)$
(thick lines), and $\nu\bar\nu$-annihilation efficiency, $q_{\nu\bar\nu}
\equiv \dot E_{\nu\bar\nu}/L_{\nu}$ (thin lines),
for different models as functions of time.}
\label{fig02}
\end{figure}
%******************************* fig02 *************************************

Since the integral rate $\dot{E}_{\nu\bar\nu}$ of energy deposition 
by neutrino-antineutrino annihilation to electron-positron pairs
($\nu\bar\nu\longrightarrow e^+e^-$) in the surroundings of the
BH-torus system scales with the product of neutrino and 
antineutrino luminosities,
\begin{equation}
\dot E_{\nu\bar\nu} \,=\, C\,L_{\nu} L_{\bar\nu}
\left( \frac{\displaystyle \langle \epsilon_{\nu}^2\rangle
\langle \epsilon_{\bar\nu}\rangle + \langle \epsilon_{\bar\nu}^2\rangle
\langle \epsilon_{\nu} \rangle}
{\displaystyle \langle \epsilon_{\nu}\rangle
\langle \epsilon_{\bar\nu}\rangle} \right) \,\, ,
\label{eq:efneu}
\end{equation}
this rate follows closely the behavior of 
$L_{\nu}$. This can be seen by a comparison of the middle panel 
of Fig.~\ref{fig01}, which shows $L_{\nu}(t)$, with the right
panel of this figure, which gives $\dot{E}_{\nu\bar\nu}(t)$. 
The factors $\ave{\epsilon_\nu}$ and $\ave{\epsilon_\nu^2}$ in
Eq.~(\ref{eq:efneu}) denote the mean energy and mean squared energy
for $\nu$ and $\bar\nu$. The factor $C$ contains weak interaction
coefficients and terms that depend on the geometry
of the neutrino-emitting torus region. The latter does not change 
much during the accretion process. 
Note that in the considered problem only $\nu_e$
and $\bar\nu_e$ need to be taken into account for computing the
annihilation rate. The contributions from muon and tau neutrinos
and antineutrinos can safely be ignored because their luminosities
are lower than those of $\nu_e$ and $\bar\nu_e$ 
by factors of several, and their annihilation cross section
is also smaller by a factor of about 5.
The general decline of $\dot{E}_{\nu\bar\nu}$ that is visible
after 15--20$\,$ms in all cases except Model ir4-64, can be
rather well fitted by a $t^{-3/2}$ power-law.

Figure~\ref{fig02} displays the conversion efficiency of rest-mass
energy to neutrinos, $q_\nu\equiv L_\nu/(\dot{M}c^2)$, as function
of time, and the corresponding efficiency for conversion of 
neutrino energy to $e^+e^-$-pairs by $\nu\bar\nu$-annihilation,
$q_{\nu\bar\nu}\equiv \dot{E}_{\nu\bar\nu}/L_{\nu}$. Values of
several per cent can be reached for both quantities in case of the
models with viscosity $\alpha = 0.1$, Models al4-64, ar1-64, and
ar2-64. 

From the results listed in Table~\ref{tab1} and Figs.~\ref{fig01}
and \ref{fig02} it is clear that only models with large disc
viscosity produce sufficiently high neutrino luminosities 
to allow for $\nu\bar\nu$-annihilation as potential energy source
of cosmological GRBs. This is apparent from
Fig.~\ref{fig03} where ray-tracing images of the neutrino emission
of the non-viscous Models r00-64, ro2-64, and ro5-64 are contrasted
by the high-viscosity Models al4-64, ar1-64, and ar2-64. The latter
models show high temperature and intense neutrino emission in a
much more extended volume around the BH. Despite of the different
impression left by the lower plots of Fig.~\ref{fig03}, there is 
only one of the investigated models where the torus is optically 
thick to neutrinos during the computed evolution, namely Model ir4-64
(not displayed in Fig.~\ref{fig03}), which has quite a massive torus
with the highest density of all models in Table~\ref{tab1}. This 
confirms that in most cases the treatment of neutrino effects by 
a trapping scheme is adequate and neutrino transport is not important.

Model ar2-64 is the most extreme case with high viscosity, high torus
mass, and direct rotation of a Kerr-BH. It combines the conditions
which are most favorable for producing GRBs. It shows the highest
torus temperatures and neutrino luminosities, yielding a total energy
deposition rate by $\nu\bar\nu$-annihilation of more than 
$10^{51}\,$erg$\,$s$^{-1}$ at 40$\,$ms when the simulations were 
stopped. At $t = 12\,$ms the value of 
$\dot{E}_{\nu\bar\nu}$ is even around $10^{52}\,$erg$\,$s$^{-1}$
(Figs.~\ref{fig01} and \ref{fig04}). Most of this energy,
however, is deposited close to the equatorial plane in a region where
the torus density is high and the rate of energy loss by neutrino 
emission is larger than the energy input by $\nu\bar\nu$-annihilation
(cf.\ the middle and right panels of Fig.~\ref{fig04}).
Therefore this energy is useless for powering relativistic outflow.
But a low-density funnel has developed along the system axis 
above the poles of the BH. Within this funnel the energy deposition by
$\nu\bar\nu$-annihilation exceeds the local reemission of neutrinos
and accounts for an energy input rate to that region of order 
$10^{50}\,$erg$\,$s$^{-1}$. With a torus lifetime of several tenths 
of a second (Table~\ref{tab1}), some $10^{49}\,$erg might thus be
suited to power a pair of ultrarelativistic axial $e^+e^-$-plasma 
jets. The involved energy is sufficiently large to make 
$\nu\bar\nu$-annihilation a viable energy source for short GRBs.

%****************************** fig03 ***************************************
\begin{figure*}
\begin{tabular}{ccc}
  \psfig{file=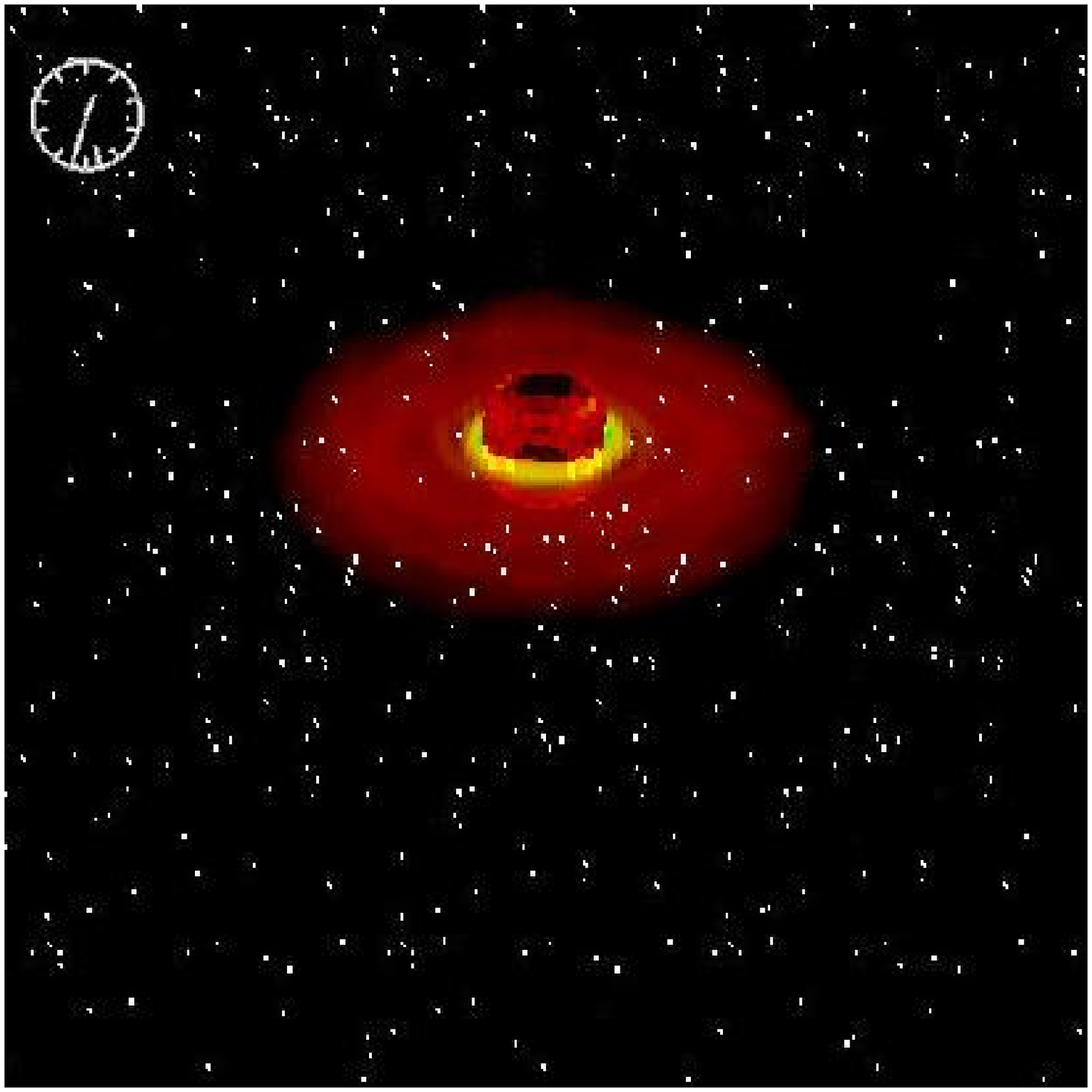,width=5.0cm} &
  \psfig{file=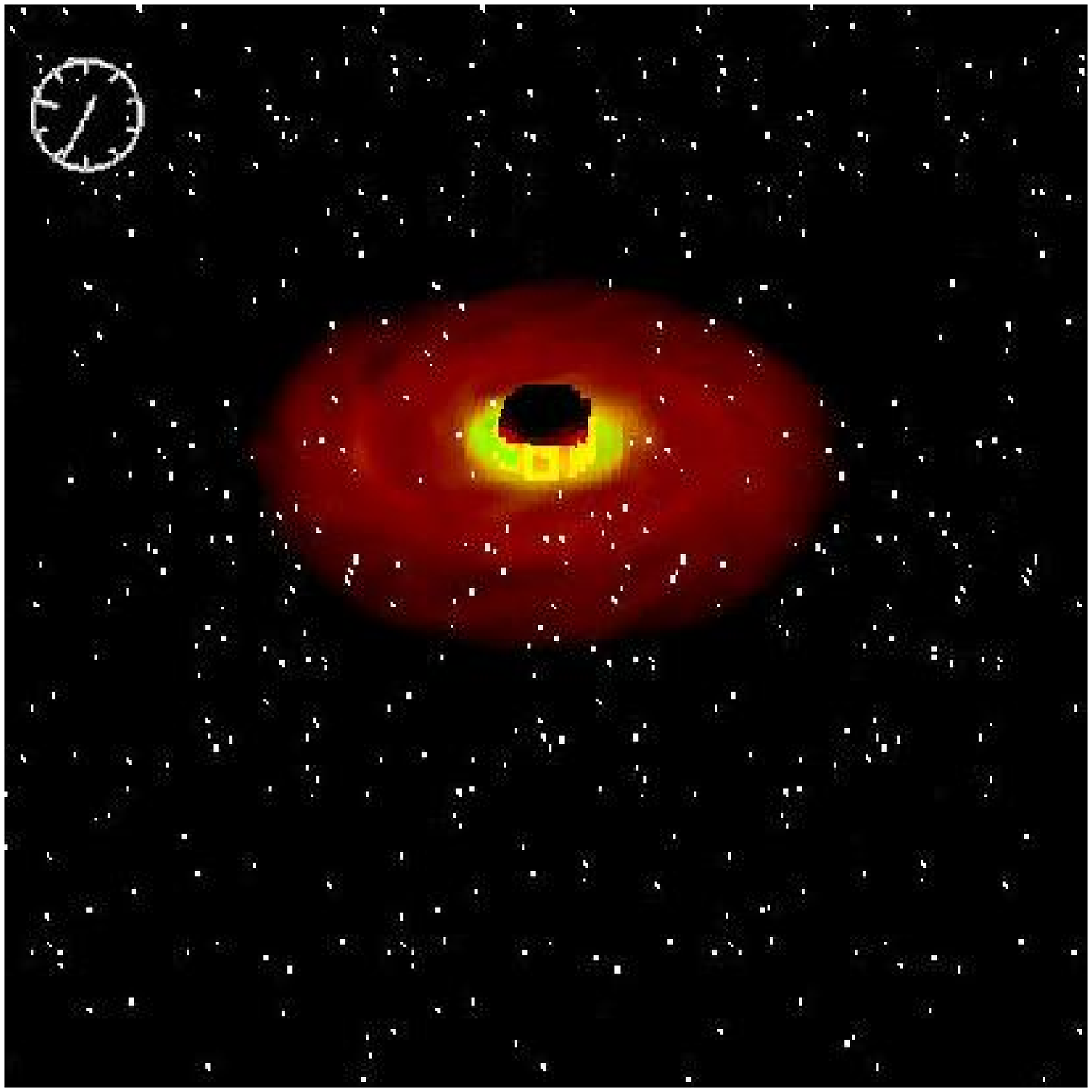,width=5.0cm} &
  \psfig{file=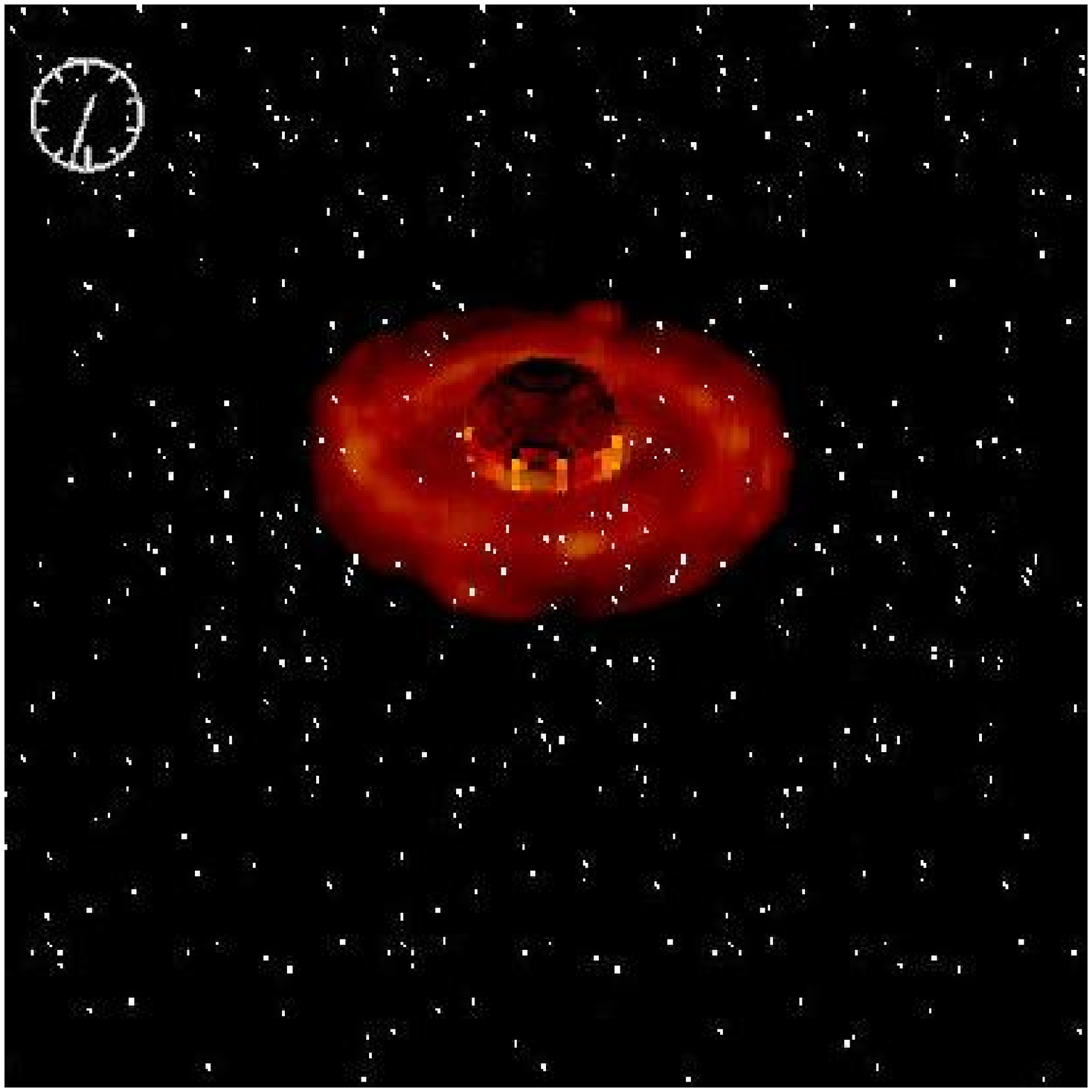,width=5.0cm} \\
  \psfig{file=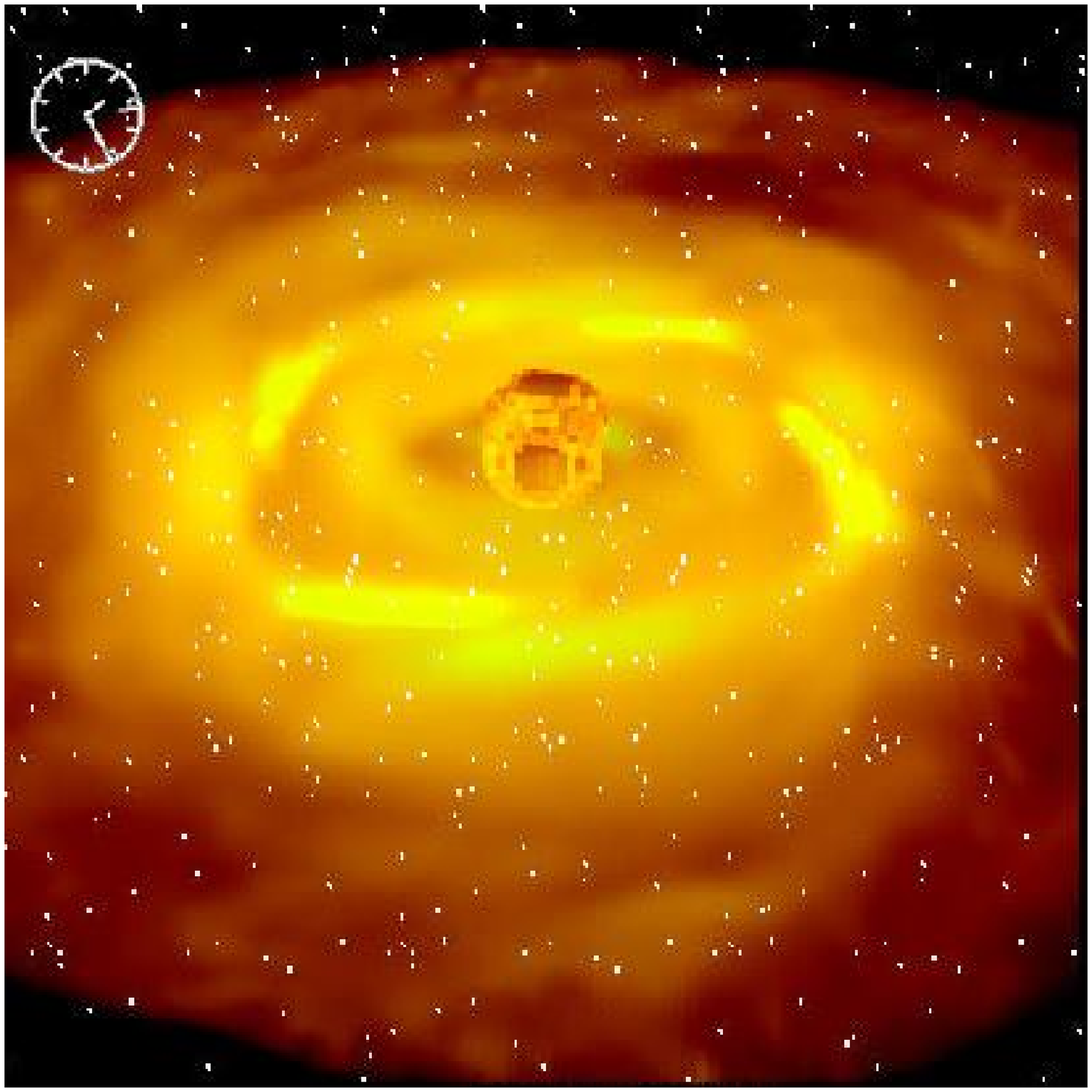,width=5.0cm} &
  \psfig{file=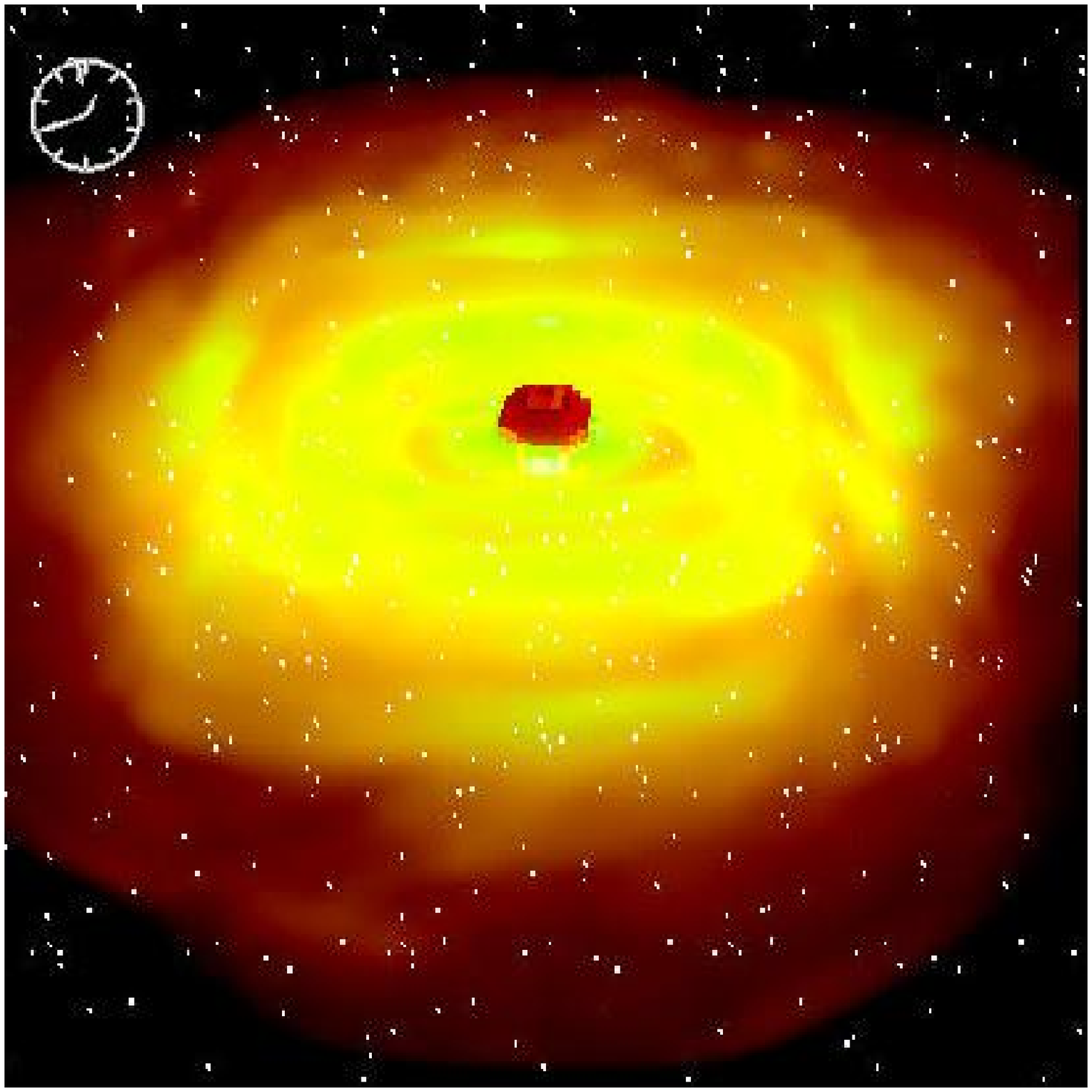,width=5.0cm} &
  \psfig{file=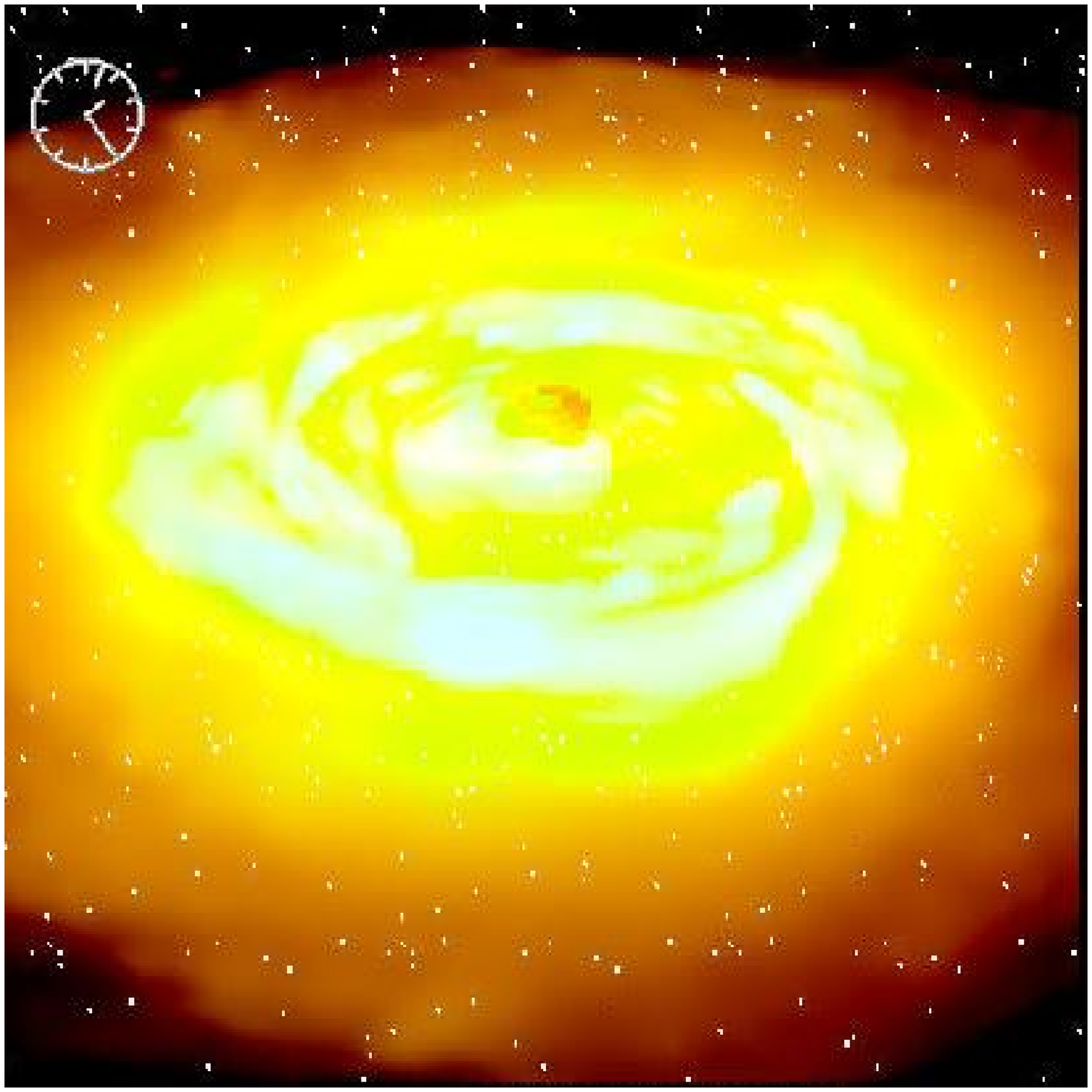,width=5.0cm}
%\vspace{0.5cm}
%\\[-2ex]
\end{tabular}
\caption[]{
Ray-tracing images of the neutrino emission of the zero-viscosity
Models r00-64, ro2-64, and ro5-64 (upper plots, from left to right)
in comparison to Models al4-64, ar1-64, and ar2-64 (lower plots,
from left to right) which were computed with a viscosity value
of $\alpha = 0.1$. The colors (ranging from red, orange, yellow,
green, blue to white) visualize the gas temperature (roughly 2, 4,
6, 8, 10, and 12$\,$MeV or higher), the brightness is a measure of
the intensity of the neutrino emission. The figures are snapshots
that correspond to times (from top left to bottom right) 5.5, 5.8,
5.5, 14.2, 7.0, and 14.0$\,$ms, respectively, after the start of the
simulations. A region with a diameter of about 500$\,$km around the
BH at the centre is displayed.
The dark spheres have a radius given by the arithmetic
mean of the event horizon and ISCO of the BH with values of, from top
left to bottom right, 24, 17, 29, 24, 17, and 17$\,$km, respectively.
The white dots symbolize background ``stars'' which are obscured by
a neutrino-opaque accretion torus. The zero-viscosity tori are
sufficiently hot and dense only very close to the BH (in particular 
in case of direct BH rotation, Model r02-64), while viscosity
leads to heating and high neutrino emission in a much more extended
volume.
}
\label{fig03}
\end{figure*}
%****************************** fig03 **************************************

%****************************** fig04 **************************************
\begin{figure*}
\begin{tabular}{ccc}
 \psfig{file=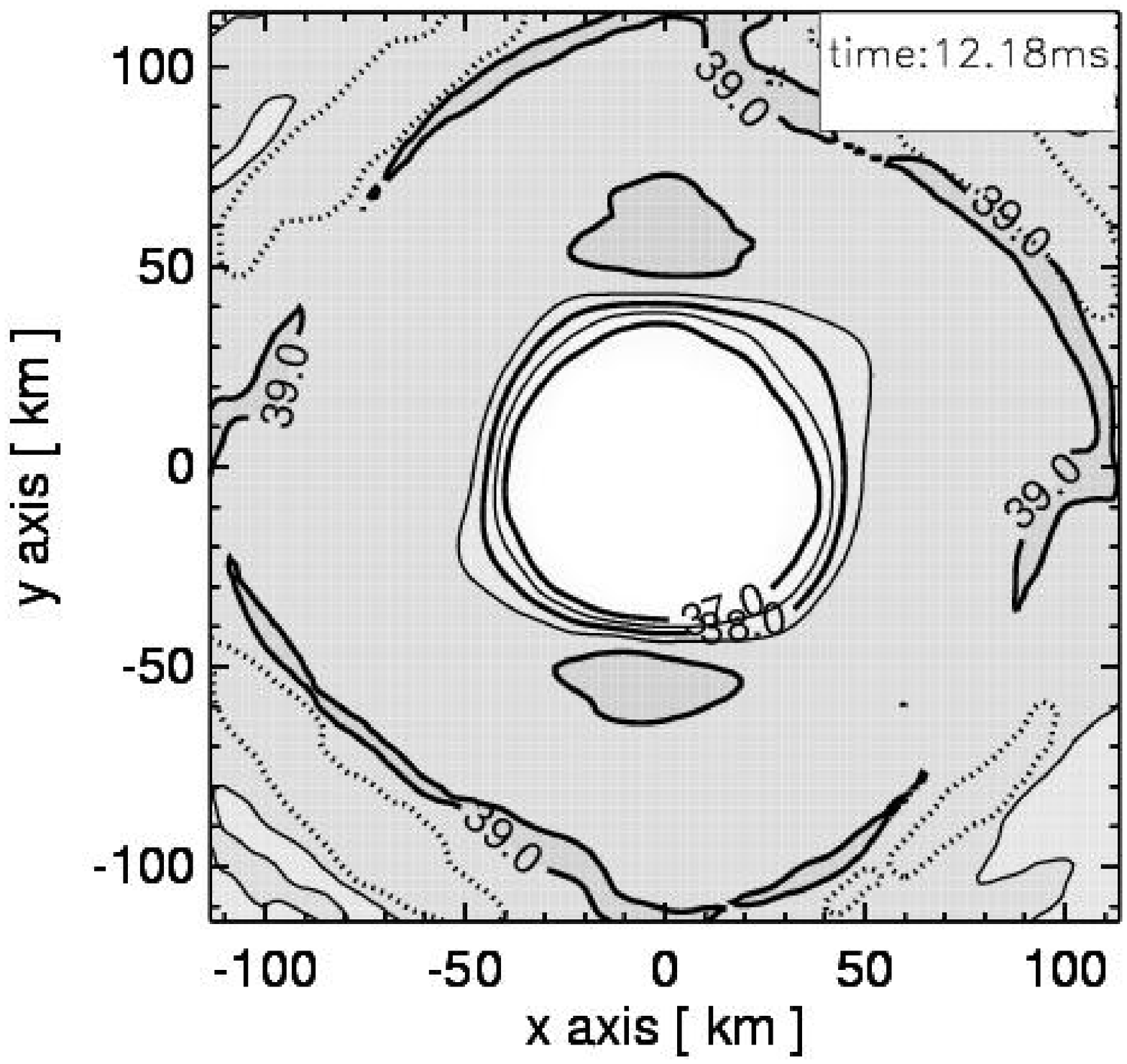,width=5.5cm,clip=} &
 \psfig{file=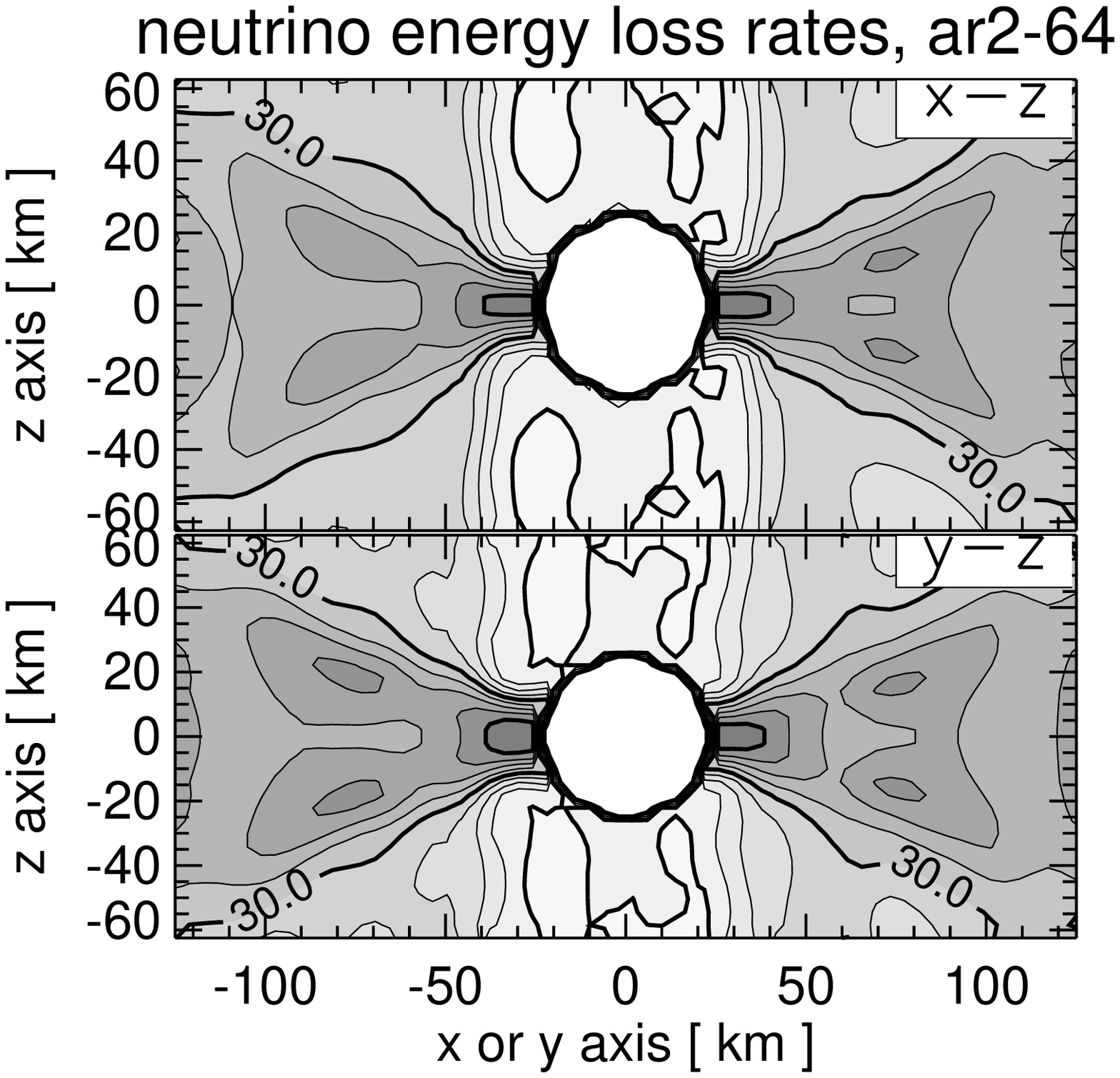,width=5.5cm,clip=} &
 \psfig{file=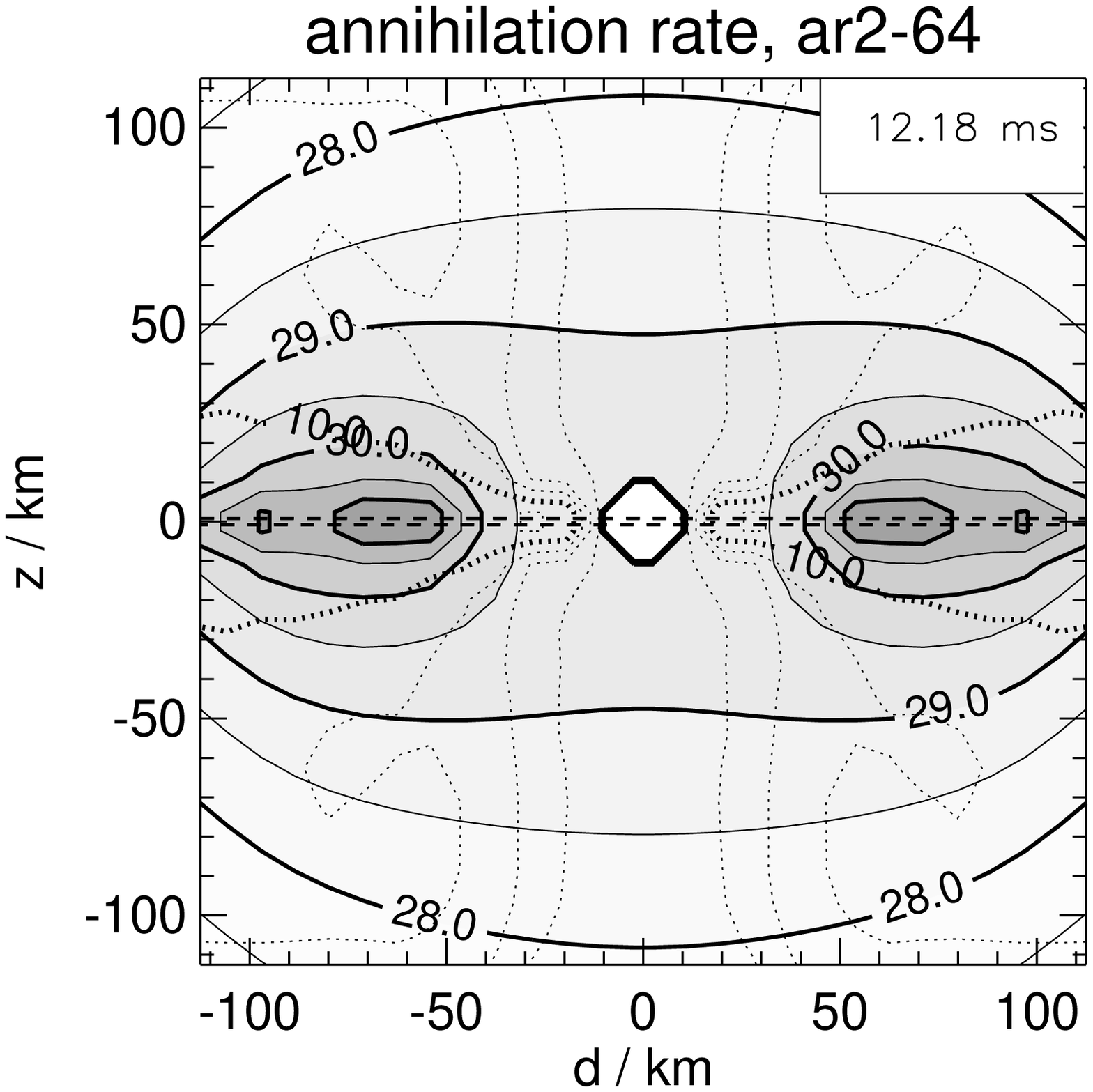,width=5.5cm,clip=}
\end{tabular}
\caption[]{
Neutrino emission and $\nu\bar\nu$-annihilation in Model~ar2-64 at time 
12.18$\,$ms. The {\em left panel} displays the surface emissivity of 
$\nu_e$ projected onto
the equatorial plane. The contours are labeled with the logarithm of
the energy loss rate in erg$\,$cm$^{-2}$s$^{-1}$. They are spaced in
steps of half a unit. The dotted lines indicate the region where
the optical depth for $\nu_e$ exceeds unity; the torus is essentially
transparent to neutrinos.
In the {\em middle panel} the energy loss rate per unit volume 
(in ${\rm erg\,cm}^{-3}{\rm s}^{-1}$) in neutrinos and antineutrinos of
all flavors is plotted in the $x$-$z$- and $y$-$z$-planes perpendicular
to the equatorial plane. The contours represent the logarithm of the
rate and are spaced in steps of half a unit. 
The {\em right panel} shows a map of the local energy deposition rate
(in erg$\,$cm$^{-3}\,$s$^{-1}$) by $\nu\bar\nu$ annihilation
in the surroundings of the BH which is indicated by the white
octagon at the centre. The
contours are logarithmically spaced with the darker grey shading
meaning higher energy deposition rate. The integral value of the
deposition rate at the displayed time is $8.5\times
10^{51}\,{\rm erg\,s}^{-1}$. The dotted contour lines represent levels
of constant values of the azimuthally averaged mass density 
(spaced logarithmically with intervals of 0.5~dex) with the
bold dotted line corresponding to $\rho=10^{10}\,{\rm g\,cm}^{-3}$.  
}
\label{fig04}
\end{figure*}
%****************************** fig04 **************************************

\section{Discussion}

We have performed three-dimensional hydrodynamic simulations
of hyperaccreting stellar mass BHs with self-gravitating, 
thick accretion tori, varying the torus mass, BH spin and 
$\alpha$-viscosity of the torus gas. Our models show that
tori which are heated by viscous dissipation of energy
get inflated by thermal pressure and are therefore unlikely to 
become optically thick to neutrinos. The neutrino luminosities
stay well below the Eddington limit $L_{\nu, {\rm Edd}} = 4\pi
G M_{\rm BH}c/\kappa_{\nu}\sim 2\times 10^{55}\,(M_{\rm BH}/4\,
M_{\odot})\,$erg$\,$s$^{-1}$. Here
$\kappa_{\nu} \sim 10^{-17}\ave{\epsilon_{\nu}^2}/(20\,{\rm MeV})^2$
is the mean neutrino opacity (e.g., Janka 2001).
The latter is mostly determined by the 
contribution of $\bar\nu_e$ which dominate the energy loss of
the torus. 
These findings are in contrast to conclusions drawn from 
more approximative modeling approaches which treated the vertical 
structure of the torus by height-averaging (Di Matteo et al.\ 2002).

Our results also suggest that the integral rate of the 
energy deposition by
$\nu\bar\nu$-annihilation drops like $t^{-3/2}$ during the
long-time, slow decay of the accretion rate that follows an 
initial, transient relaxation phase of 10--20$\,$ms with very 
high mass accretion rates and neutrino emission. This temporal 
decrease is less steep than obtained in previous simulations 
with azimuthal symmetry which used a simple ideal gas equation
state and made the assumption
that all the dissipated energy is radiated away in
neutrinos (Lee \& Ramirez-Ruiz 2002).

We have shown that $\nu\bar\nu$-annihilation in the low-density
funnel above the poles of the BH can deposit energy at a rate of
$\sim 10^{50}\,$erg$\,$s$^{-1}$, accounting for a total energy 
release of some $10^{49}\,$erg in case of sufficiently large
torus mass
($\ga\,$0.1$\,M_{\odot}$), high disk viscosity ($\alpha\sim 0.1$)
and BH rotation with spin parameter $a\sim 0.6$. 
These conditions are expected to be 
generically produced by BH+NS mergers and most likely also
by NS+NS mergers (Janka et al.\ 1999). The main effect of direct
BH rotation is an increase of the lifetime of the torus.
The energy release by
$\nu\bar\nu$-annihilation satisfies the energy requirements
of short GRBs in case of a moderate amount of collimation of the
ultrarelativistic bipolar outflows into a fraction 
$f = 2\Delta\Omega/4\pi$ of a few per cent of the sky (e.g.,
Rosswog \& Ramirez-Ruiz 2003). Our results for the post-merging
accretion of a remnant BH-torus system are therefore more 
optimistic than the estimates based on NS+NS merger simulations
by Rosswog \& Ramirez-Ruiz (2002). Future simulations, however,
will have to show whether the deposited thermal energy can
be efficiently converted to axial outflows from BH-torus systems
that are sufficiently luminous and collimated to account for 
short GRBs (Aloy, Janka \& M\"uller, in preparation).

\section*{Acknowledgments}
SS highly appreciated continuing encouragement from S.F. Gull 
and is grateful to G.I. Ogilvie for valuable discussions.
We thank the anonymous referee for suggestions which helped to
improve our paper.
SS acknowledges support from the Particle Physics and Astronomy Research 
Council (PPARC), HTJ from the Sonderforschungsbereich 375 
``Astro-Teilchenphysik'' und the Sonderforschungsbereich-Transregio 7
``Gravi\-ta\-tions\-wel\-len\-astronomie'' of 
the Deutsche Forschungsgemeinschaft.
Parts of the simulations were performed at the UK Astrophysical Fluids 
Facility (UKAFF) and the Edinburgh Parallel Computing Centre (EPCC) of 
the University of Edinburgh.

{}


\begin{thebibliography}{}

\bibitem[1996]{art96}		Artemova I.V., Bj\"ornsson G., 
				Novikov I.D., 1996, ApJ, 461, 565
\bibitem[1990]{bla90}   	Blanchet L., Damour T., Sch\"afer G., 
				1990, MNRAS, 242, 289
\bibitem[1984]{col84}   	Colella P., Woodward P.R., 1984, 
				JCP, 54, 174
\bibitem[2002]{di02}		Di Matteo T., Perna R., Narayan R., 2002, 
				ApJ, 579, 706
\bibitem[1989]{ei89}            Eichler D., Livio M., Piran T., 
                                Schramm D.N., 1989, Nature, 340, 126
\bibitem[2002]{ko02}            Kohri K., Mineshige S., 2002, ApJ, 577, 311
\bibitem[1991]{lat91}  		Lattimer J.M., Swesty F.D., 1991, 
				Nucl.~Phys.~A, 535, 331
\bibitem[2001]{lee01}           Lee W.H., 2001, MNRAS, 328, 583
\bibitem[2002]{lee02}		Lee W.H., Ramirez-Ruiz E., 2002, 
				ApJ, 577, 893
\bibitem[2001]{jan01}           Janka H.-Th., 2001, A\&A, 368, 527
\bibitem[1999]{jan99}		Janka H.-Th., Eberl T., Ruffert M., 
				Fryer C.L., 1999, ApJ, 527, L39
\bibitem[1997]{mes97}   	M\'esz\'aros P., Rees M.J., 1997, 
				ApJ, 482, L29
\bibitem[1992]{nar92}           Narayan R., Paczy\'nski B., Piran T.,
                                1992, ApJ, 395, L83
\bibitem[2001]{nar01}		Narayan R., Piran T., Kumar P., 2001, 
				ApJ, 557, 949
\bibitem[2004]{oe04}		Oechslin R., Ury\={u} K., Poghosyan G., 
				Thielemann F.-K., 2004, MNRAS, in press; 
                                astro-ph/0401083
\bibitem[1980]{pac80}   	Paczy\'nski B., Wiita P.J., 1980, 
				A\&A, 88, 23
\bibitem[1999]{pop99}  		Popham R., Woosley S.E., Fryer C., 1999, 
				ApJ, 518, 356
\bibitem[2003]{roslie03}        Rosswog S., Liebend\"orfer M., 2003,
                                MNRAS, 342, 673
\bibitem[2002]{ros02}		Rosswog S., Ramirez-Ruiz E., 2002, 
                                MNRAS, 336, L7 
\bibitem[2003]{ros03}		Rosswog S., Ramirez-Ruiz E., 2003, 
                                MNRAS, 343, L36
\bibitem[1996]{ruffert1} 	Ruffert M., Janka H.-Th., Sch\"{a}fer G.,
  				1996, A\&A, 311, 532
\bibitem[1997]{ruffert2} 	Ruffert M., Janka H.-Th., Takahashi K., 
				Sch\"{a}fer G., 1997, A\&A, 319, 122
\bibitem[1999]{ruffert} 	Ruffert M., Janka H.-Th., 1999, 
				A\&A, 344, 573
\bibitem[2001]{ruffert3} 	Ruffert M., Janka H.-Th., 2001, 
				A\&A, 380, 544
\bibitem[1973]{sha73} 		Shakura N.I., Sunyaev R.A., 1973,
				A\&A, 24, 337
\bibitem[2000]{shi00}		Shibata M., Ury\={u} K., 2000, 
                                PRD, 61, 064001
\bibitem[1993]{woo93}    	Woosley S.E., 1993, ApJ, 405, 273

\end{thebibliography}
\end{document}